\documentclass[aps,prd,nofootinbib,amsmath,amssymb,twocolumn,superscriptaddress,10pt]{revtex4}

\pdfoutput=1

\usepackage{graphicx}
\usepackage{dcolumn}
\usepackage{bm}
\usepackage{amssymb}
\usepackage{latexsym}
\usepackage{booktabs}
\usepackage{amsmath}
\usepackage{multirow}
\usepackage[colorlinks=true, linkcolor=red, citecolor=blue]{hyperref}

\newcommand{\be}{\begin{equation}}
\newcommand{\ee}{\end{equation}}
\newcommand{\bq}{\begin{eqnarray}}
\newcommand{\eq}{\end{eqnarray}}

\usepackage[usenames,dvipsnames]{xcolor}

\DeclareMathAlphabet\mathbfcal{OMS}{cmsy}{b}{n}
\definecolor{darkgreen}{cmyk}{0.85,0.2,1.00,0.2}
\definecolor{purple}{cmyk}{0.5,1.0,0,0}

\def\barray{\begin{array}}
\def\earray{\end{array}}
\def\be{\begin{equation}}
\def\ee{\end{equation}}
\def\ben{\begin{equation} \nonumber}
\def\een{\end{equation}}
\def\ban{\begin{eqnarray*}}
\def\ean{\end{eqnarray*}}
\def\ba{\begin{eqnarray}}
\def\ea{\end{eqnarray}}

\def\({\left(}
\def\){\right)}

\begin{document}

\title{Reexploration of interacting holographic dark energy model: Cases of interaction term excluding the Hubble parameter}
\author{Hai-Li Li}
\affiliation{Department of Physics, College of Sciences,
Northeastern University, Shenyang 110004, China}
\author{Jing-Fei Zhang}
\affiliation{Department of Physics, College of Sciences,
Northeastern University, Shenyang 110004, China}
\author{Lu Feng}
\affiliation{Department of Physics, College of Sciences,
Northeastern University, Shenyang 110004, China}
\author{Xin Zhang}
\affiliation{Department of Physics, College of Sciences,
Northeastern University, Shenyang 110004, China}
\affiliation{Center for High Energy Physics, Peking University, Beijing 100080, China}

\begin{abstract}

In this paper, we make a deep analysis for the five typical interacting holographic dark energy models with the interaction terms $Q=3\beta H_{0}\rho_{\rm{de}}$, $Q=3\beta H_{0}\rho_{\rm{c}}$, $Q=3\beta H_{0}(\rho_{\rm{de}}+\rho_{\rm c})$, $Q=3\beta H_{0}\sqrt{\rho_{\rm{de}}\rho_{\rm c}}$, and $Q=3\beta H_{0}\frac{\rho_{\rm{de}}\rho_{c}}{\rho_{\rm{de}}+\rho_{\rm c}}$, respectively. We obtain observational constraints on these models by using the type Ia supernova data (the Joint Light-curve Analysis sample), the cosmic microwave background data (Planck 2015 distance priors), the baryon acoustic oscillations data, and the direct measurement of the Hubble constant. We find that the values of $\chi_{\rm min}^2$ for all the five models are almost equal (around~699), indicating that the current observational data equally favor these IHDE models. In addition, a comparison with the cases of interaction term involving the Hubble parameter $H$ is also made.

\end{abstract}
 \maketitle

\section{Introduction}\label{sec:intro}

The accelerated expansion of the universe has been discovered by the type Ia supernova observations \cite{Riess:1998cb,Perlmutter:1998np} and further confirmed by various other cosmological observations \cite{Spergel:2003cb,Bennett:2003bz,Tegmark:2003ud,Abazajian:2004aja}. Dark energy that has negative pressure has been proposed to explain the phenomenon of cosmic acceleration \cite{Sahni:2006pa,Bamba:2012cp,Weinberg:1988cp,Peebles:2002gy,Copeland:2006wr,Frieman:2008sn,Sahni:2008zz,Li:2011sd,Kamionkowski:2007wv}. In the present universe, dark energy contributes about 70\% of the cosmic energy density, and thus it is now dominating the evolution of the universe. The study of dark energy has become one of the most important issues in theoretical physics and modern cosmology. Although enormous efforts have been made to investigate dark energy, its nature is still in the dark.

The primary candidate of dark energy is the so-called ``cosmological constant'' (denoted as $\Lambda$), which is equivalent to the density of vacuum energy and thus is a constant in space and time. The cosmological constant $\Lambda$ has a constant equation-of-state parameter (EoS) $w_{\Lambda}\equiv p_\Lambda/\rho_\Lambda=-1$. The cosmological model with $\Lambda$ and cold dark matter (CDM) is called the $\Lambda$CDM model, which can explain the current various cosmological observations quite well \cite{Ade:2015xua}. However, the cosmological constant always suffers from serious theoretical challenges, i.e., the so-called ``fine-tuning'' and ``cosmic coincidence'' puzzles \cite{Sahni:1999gb,Bean:2005ru}. The value of the vacuum energy density calculated by quantum field theory is higher than the fit value of the cosmological constant by cosmological observations by about the 120 orders of magnitude, and so a bare cosmological constant needs to be introduced to make an offset, leading to the fine-tuning problem. The coincidence problem asks why the densities of vacuum energy and matter are in the same order today, although their evolutionary histories differ enormously. These two puzzles have been frustrating the $\Lambda$CDM cosmology in theoretical aspect.

Actually, there are many other candidates for dark energy, for which the vast majority believes that dark energy has dynamics, often realized by some scalar field \cite{Steinhardt:1999nw,Zlatev:1998tr,Liddle:1998xm,Caldwell:1997ii,Turner:1998ex,Frieman:1995pm,Guo:2004fq,Zhao:2006mp,Wei:2005nw,Cai:2007gs,Cai:2009zp}. However, more theoretically, it is believed that the dark energy problems are closely related to the theory of quantum gravity in nature. It is actually obvious that $\Lambda$ has a quantum origin and at the same time it yields repulsive gravity leading the current universe to accelerate. Therefore, it is of great interest to explore the nature of dark energy from the perspective of quantum gravity. In the current circumstance that we have no a complete theory of quantum gravity, we have to appeal to the holographic principle of quantum gravity for an effective theory of dark energy.

Together with the effective quantum field theory, the holographic principle leads to a model of dark energy, named ``holographic dark energy'' (HDE) model, which can solve the two theoretical puzzles of cosmological constant at the same time to some extent \cite{Li:2004rb} and can explain the current cosmological observations well (see Ref.~\cite{Wang:2016och} for a recent review). By considering the holographic principle, it is required that in a spatial region there is an upper limit for the number of degrees of freedom involved in it, due to the gravitational effects of them (the condition of black hole formation sets such a bound) \cite{Cohen:1998zx}. That is to say, in this theory, the infrared (IR) cutoff (with length scale $L$) is related to the ultraviolet (UV) cutoff (with energy scale $k_{\rm max}$). Recall that for the vacuum energy density we have the evaluation $\rho_{\rm vac}\simeq k_{\rm max}^4/(16\pi^2)$. Thus, one finds that by such a theoretical consideration the density of dark energy can be decided by some IR cutoff length scale of the universe. The holographic reasoning gives the density of dark energy of the form \cite{Li:2004rb}
\begin{equation}\label{1.1}
 \rho_{\rm de}=3c^2 M^2_{\rm pl} L^{-2},
\end{equation}
where $M_{\rm pl}=\frac{1}{\sqrt{8\pi G}}$ is the reduced Planck mass and $c$ is a dimensionless constant characterizing ambiguous factors in the effective theory. In the HDE model, Li \cite{Li:2004rb} argues that $L$ should be chosen as the future event horizon of the universe, which can lead to a cosmic acceleration. Namely, in the HDE model, we have
\begin{equation}\label{1.2}
L=a(t)\int_{t}^{\infty}\frac{dt'}{a(t')}=a\int_{a}^{\infty}\frac{da'}{Ha'^2}.
\end{equation}
The HDE model has been widely studied in depth \cite{Huang:2004wt,Wang:2004nqa,Huang:2004mx,Chang:2005ph,Nojiri:2005pu,Zhang:2006av,Zhang:2006qu,Zhang:2007sh,Zhang:2007es,Zhang:2007an,Li:2008zq,Ma:2007av,Li:2009bn,Zhang:2009xj,Li:2013dha,Zhang:2015rha,Cui:2015oda,delCampo:2011jp,Landim:2015hqa,Zhao:2017urm}. There are also some variants of this kind \cite{Wei:2007ty,Gao:2007ep,Cai:2008nk,Zhang:2009un,He:2016rvp,Liao:2012qk,Zhang:2010im,Zhang:2012pr,Li:2012xm,Wu:2008jt}. In the HDE model, it is found that the parameter $c$ solely determines the evolution of dark energy, by solving a differential equation (see, e.g., Ref.~\cite{Li:2004rb}). When $c>1$, the dark energy has $w>-1$ (in the case of $c=1$, $w$ will eventually evolve to get $-1$); when $c<1$, the EoS of dark energy $w$ will cross the phantom divide $-1$ from $w\rightarrow -1/3$ to $w\rightarrow -1/3-2/(3c)$ \cite{Zhang:2005hs}. The cosmological constraints show that $c$ is around 0.7 (see, e.g., Refs.~\cite{Zhang:2014ija,Ma:2007pd,Wang:2012uf}). A recent work \cite{Xu:2016grp} on comparing popular dark energy models shows that the HDE model is still a competitive model in the aspect of fitting the current cosmological observations.

On the other hand, there might be some direct interaction between dark energy and dark matter, which is capable of helping resolve (or alleviate) the coincidence problem of dark energy \cite{Zhang:2005rj,Amendola:1999er,Cai:2004dk}. Therefore, addition to probing the dynamics of dark energy, another important mission for the investigation of dark energy is to detect such a ``fifth force'' between dark energy and dark matter by accurate cosmological observations. The interacting dark energy models have been widely studied \cite{Amendola:2001rc,Comelli:2003cv,Zimdahl:2005bk,Wang:2006qw,Guo:2007zk,Bertolami:2007zm,Boehmer:2008av,He:2008tn,He:2009mz,He:2009pd,Koyama:2009gd,Xia:2009zzb,He:2010im,Chen:2011rz,Clemson:2011an,Wang:2013qy,Fu:2011ab,Wang:2014oga,Yin:2015pqa,Murgia:2016ccp,Sola:2016ecz,Sola:2016jky,Sola:2016zeg,Sola:2017jbl,Pourtsidou:2016ico,Costa:2016tpb,Xia:2016vnp,vandeBruck:2016hpz,Kumar:2016zpg,Kumar:2017dnp,Santos:2017bqm,Valiviita:2008iv,Guo:2017hea,Zhang:2017ize,Zhang:2004gc,Cai:2009ht,Li:2011ga,Wei:2010cs,Zhang:2012sya}. The interacting models in the framework of holographic dark energy have also been deeply explored (see, e.g., Refs.~\cite{Li:2009zs,Zhang:2012uu,Feng:2016djj}).

In interacting models of dark energy, one considers that there is an energy transfer between dark energy and dark matter in the background universe (and in a perturbed universe there is also a momentum transfer between them). In a concrete model, a form of the energy (density) transfer rate (denoted as $Q$) should be assumed. Usually, consulting from the theories of nuclear decay and inflationary reheating, the form of $Q$ is assumed to be proportional to the density of dark energy or dark matter, i.e., $Q=3\beta H\rho_{\rm de}$ or $Q=3\beta H\rho_{\rm c}$, where $\rho_{\rm de}$ and $\rho_{\rm c}$ are the densities of dark energy and cold dark matter, respectively, $H$ is the Hubble parameter, and $\beta$ denotes the dimensionless coupling between dark energy and dark matter. Note that here $3H$ appears only for mathematical convenience. In our recent work \cite{Feng:2016djj}, Feng and Zhang explored the interacting models in the framework of holographic dark energy and made a comparison for five interacting cases ($Q=3\beta H\rho_{\rm{de}}$, $Q=3\beta H\rho_{\rm{c}}$, $Q=3\beta H(\rho_{\rm{de}}+\rho_{\rm c})$, $Q=3\beta H\sqrt{\rho_{\rm{de}}\rho_{\rm c}}$, and $Q=3\beta H\frac{\rho_{\rm{de}}\rho_{c}}{\rho_{\rm{de}}+\rho_{\rm c}}$), according to the constraint results of current observations.

However, in the research area of interacting dark energy, there is another perspective that $Q$ should not involve the Hubble parameter $H$ because the local interaction should not depend on the global expansion of the universe \cite{Valiviita:2008iv}. According to this perspective, one should write down the form of $Q$ as, e.g., $Q=3\beta H_0\rho_{\rm de}$ or $Q=3\beta H_0\rho_{\rm c}$, where the appearance of $H_0$ is only for a dimensional consideration. In this paper, we will revisit the exploration of interacting holographic dark energy models by adopting this perspective. We will consider the five cases with $Q=3\beta H_{0}\rho_{\rm{de}}$, $Q=3\beta H_{0}\rho_{\rm{c}}$, $Q=3\beta H_{0}(\rho_{\rm{de}}+\rho_{\rm c})$, $Q=3\beta H_{0}\sqrt{\rho_{\rm{de}}\rho_{\rm c}}$, and $Q=3\beta H_{0}\frac{\rho_{\rm{de}}\rho_{c}}{\rho_{\rm{de}}+\rho_{\rm c}}$. We constrain the models by using the current cosmological observations, and we report the results and make an analysis for them.

This paper is organized as follows. In Sec.~\ref{sec2}, we briefly describe the interacting holographic dark energy model. In Sec.~\ref{sec3}, we present the analysis method and the observational data used in this paper. In Sec.~\ref{sec4}, we report the constraint results and make a deep discussion for them. Conclusion is given in Sec.~\ref{sec5}.

\section{The interacting model of holographic dark energy}\label{sec2}

In this section, we briefly derive the equations describing the interacting holographic dark energy (IHDE) model for a flat cosmology.

In the context of flat Friedmann-Roberston-Walker universe, the Friedmann equation can be written as
\begin{equation}\label{2.1}
3M^2_{\rm{pl}} H^2=\rho_{\rm c}+\rho_{\rm b}+\rho_{\rm r}+\rho_{\rm{de}},
\end{equation}
where $3M^2_{\rm{pl}} H^2$ is the critical density of the universe, $\rho_{\rm c}$, $\rho_{\rm b}$, $\rho_{\rm r}$, and $\rho_{\rm{de}}$ represent the energy densities of cold dark matter, baryon, radiation, and dark energy, respectively. For convenience, we define the fractional energy densities of various components as
\begin{equation}\label{2.2}
\begin{aligned}
\Omega_{\rm c}=\frac{\rho_{\rm c}}{3M^2_{\rm{pl}} H^2},~~\Omega_{\rm b}=\frac{\rho_{\rm b}}{3M^2_{\rm{pl}} H^2},\\
\Omega_{\rm r}=\frac{\rho_{\rm r}}{3M^2_{\rm{pl}} H^2},~~\Omega_{\rm de}=\frac{\rho_{\rm de}}{3M^2_{\rm{pl}} H^2}.
\end{aligned}
\end{equation}
By definition, we have
\begin{equation}\label{2.3}
\Omega_{\rm{c}}+\Omega_{\rm{b}}+\Omega_{\rm r}+\Omega_{\rm{de}}=1.
\end{equation}

In the IHDE model, there is some direct, non-gravitational interaction between holographic dark energy and dark matter, and thus we have the following continuity equations for the various components:
\begin{equation}\label{2.4}
\dot{\rho}_{\rm c}+3H\rho_{\rm c}=Q,
\end{equation}
\begin{equation}\label{2.5}
\dot{\rho}_{\rm de}+3H(\rho_{\rm de}+p_{\rm de})=-Q,
\end{equation}
\begin{equation}\label{2.6}
\dot{\rho_{\rm b}}+3H\rho_{\rm b}=0,
\end{equation}
\begin{equation}\label{2.7}
\dot{\rho_{\rm r}}+4H\rho_{\rm r}=0,
\end{equation}
where $Q$ is the phenomenological interaction term \cite{Zhang:2005rg,Zhang:2007uh,Zhang:2009qa,Li:2010ak,Zhang:2013lea,Li:2013bya,Li:2014eha,Li:2014cee,Geng:2015ara,Li:2015vla}, denoting the energy transfer rate between dark energy and dark matter. In this paper, we consider the following five cases in the IHDE model:
\begin{equation}\label{2.8}
Q_1=3\beta H_{0}\rho_{\rm{de}},
\end{equation}
\begin{equation}\label{2.9}
Q_2=3\beta H_{0}\rho_{\rm c},
\end{equation}
\begin{equation}\label{2.10}
Q_3=3\beta H_{0}(\rho_{\rm{de}}+\rho_{\rm c}),
\end{equation}
\begin{equation}\label{2.11}
Q_4=3\beta H_{0}\sqrt{\rho_{\rm{de}}\rho_{\rm c}},
\end{equation}
\begin{equation}\label{2.12}
Q_5=3\beta H_{0}\frac{\rho_{\rm de}\rho_{\rm c}}{\rho_{\rm de}+\rho_{\rm c}}.
\end{equation}
As has been mentioned above, $\beta$ is a dimensionless coupling parameter describing the strength of interaction between dark energy and dark matter.

Combining Eqs. (\ref{2.1}) and (\ref{2.3})--(\ref{2.7}), we obtain
\begin{equation}\label{2.13}
p_{\rm de}=-\frac{2}{3}\frac{\dot{H}}{H^2}\rho_{\rm c}-\rho_{\rm c}-\frac{1}{3}\rho_{\rm r}.
\end{equation}
Substituting Eq. (\ref{2.13}) into Eq. (\ref{2.5}), we obtain
\begin{equation}\label{2.14}
2\frac{\dot{H}}{H}(\Omega_{\rm{de}}-1)+\dot{\Omega}_{\rm{de}}+H(3\Omega_{\rm{de}}+\Omega_{\rm I}-3-\Omega_{\rm r})=0.
\end{equation}
Here, for convenience, following Ref.~\cite{Zhang:2012uu} we define
\begin{equation}\label{2.15}
 \Omega_{\rm I}=\frac{Q}{3M_{\rm{pl}}^2H^3}.
\end{equation}

From Eq.~(\ref{1.1}) (i.e., the definition of density of holographic dark energy), we can get a relation,
\begin{equation}\label{2.16}
L=\frac{c}{H\sqrt{\Omega_{\rm{de}}}}.
\end{equation}
We now write the IR cut-off length scale $L$ as the form
\begin{equation}\label{2.17}
 L=ar(t).
\end{equation}
Combining Eqs. (\ref{2.16}) and (\ref{2.17}), we get
\begin{equation}\label{2.18}
r(t)=\frac{L}{a}=\frac{c}{Ha\sqrt{\Omega_{\rm{de}}}}.
\end{equation}
Combining Eqs. (\ref{1.2}) and (\ref{2.18}), we have the relation
\begin{equation}\label{2.19}
\int_{t}^{\infty}\frac{dt'}{a(t')}=\frac{c}{Ha\sqrt{\Omega_{\rm{de}}}}.
\end{equation}
Taking derivative of Eq. (\ref{2.19}) with respect to $t$, we can get the equation
\begin{equation}\label{2.20}
\frac{\dot{\Omega}_{\rm{de}}}{2\Omega_{\rm{de}}}+H+\frac{\dot{H}}{H}=\frac{H}{c}\sqrt{\Omega_{\rm{de}}}.
\end{equation}

Combining Eqs. (\ref{2.14}) and (\ref{2.20}), we get the following two differential equations governing the dynamical evolution of dark energy in the IHDE model for a flat cosmology,
\begin{equation}\label{2.21}
\frac{1}{E}\frac{dE}{dz}=-\frac{\Omega_{\rm{de}}}{1+z}\left(\frac{1}{c}\sqrt{\Omega_{\rm{de}}}+\frac{1}{2}+\frac{\Omega_{\rm I}-3-\Omega_{\rm r}}{2\Omega_{\rm{de}}}\right),
\end{equation}
\begin{equation}\label{2.22}
\frac{d\Omega_{\rm{de}}}{dz}=-\frac{2(1-\Omega_{\rm{de}})\Omega_{\rm{de}}}{1+z}\left(\frac{1}{c}\sqrt{\Omega_{\rm{ de}}}+\frac{1}{2}+\frac{\Omega_{\rm I}-\Omega_{\rm r}}{2(1-\Omega_{\rm{de}})}\right),
\end{equation}
where $E(z)=H(z)/H_0$ is the dimensionless Hubble expansion rate, $\Omega_{\rm{de}}(z)$ is the fractional density of dark energy, and $\Omega_{\rm{r}}(z)=\Omega_{\rm{r0}}(1+z)^4/E(z)^2$ is the fractional density of radiation. Here we have $\Omega_{ \rm{r0}}=\Omega_{\rm{m0}} / (1+z_{\rm eq})$, where $\Omega_{\rm m0}=\Omega_{\rm c0}+\Omega_{\rm b0}$ and $z_{\rm eq}=2.5\times 10^4 \Omega_{\rm{m0}} h^2 (T_{\rm cmb}/2.7\,{\rm K})^{-4}$, with $T_{\rm cmb}=2.7255\,{\rm K}$. The initial conditions of these equations are $E_0=1$ and $\Omega_{\rm{de0}}=1-\Omega_{\rm{m0}}-\Omega_{\rm{r0}}$ at $z=0$. 

In this paper, for convenience, we occasionally call the cases with $Q_1$--$Q_5$ [described by Eqs.~(\ref{2.8})--(\ref{2.12})] the IHDE1--IHDE5 models, respectively.

\section{Method and data}\label{sec3}

In a flat universe, the IHDE models have four free parameters, $c$, $h$, $\Omega_{m0}$, and $\beta$. We will use the current observational data to constrain the models.

We use the $\chi^2$ statistic to estimate the model parameters. The form of $\chi^2$ function is as follows,
\begin{equation}\label{3.1}
\chi^{2}_{\xi}=\frac{(\xi_{\rm th}-\xi_{\rm obs})^{2}}{\sigma^{2}_{\xi}},
\end{equation}
where $\xi_{\rm th}$ is the theoretically predicted value for the observable $\xi$, $\xi_{\rm obs}$ is the corresponding experimentally measured value, and $\sigma_{\xi}$ is the standard deviation.
The total $\chi^2$ is the sum of all $\chi^2_\xi$,
\begin{equation}\label{3.2}
\chi^2=\sum\limits_{\xi} \chi^2_\xi.
\end{equation}

The observational data we use in this paper include the type Ia supernova (SN) data, the cosmic microwave background (CMB) anisotropy data, the baryon acoustic oscillation (BAO) data, and the direct measurement of the Hubble constant $H_{0}$. Thus the total $\chi^2$ function is
\begin{equation}\label{3.3}
  \chi^2=\chi^2_{\rm SN}+\chi^2_{\rm CMB}+\chi^2_{\rm BAO}+\chi^2_{H_0}.
\end{equation}

Since these IHDE models have the same parameter number, we can direct compare them with their $\chi^2$ values. But when we compare them with the $\Lambda$CDM model and the HDE model (without interaction), the $\chi^2$ comparison becomes unfair because their parameter numbers are different. We thus employ the Akaike information criterion (AIC) \cite{AIC1974} and the Bayesian information criterion (BIC) \cite{BIC1978} to do the model comparison in this situation. By definition, we have ${\rm AIC}=-2\ln{\mathcal{L}_{\rm{max}}}+2k$ and ${\rm BIC}=-2\ln{\mathcal{L}_{\rm{max}}}+k\ln{N}$, where $k$ is the number of parameters, and $N$ is the number of data points. Since we wish to measure the difference between models, we are more interested in the relative values of them. In this work, we choose the $\Lambda$CDM as a reference model, and then calculate $\Delta {\rm AIC}=\Delta\chi^2_{\rm{min}}+2\Delta k$ and $\Delta{\rm BIC}=\Delta\chi^2_{\rm{min}}+\Delta k\ln N$. A model with a lower value of AIC or BIC is believed to be more favored by data.

\subsection{Type Ia supernovae}

We use the JLA compilation of type Ia supernovae \cite{Betoule:2014frx} in this work. It is from a joint analysis of type Ia supernova observations. The JLA compilation consists of 740 Ia supernovae data points, obtained by the SDSS-II and SNLS collaborations. The distance modulus of a SN Ia is
\begin{equation}
\hat{\mu}=m^{\ast}_{\rm{B}}-(M_{\rm{B}}-\alpha \times X_{1}+\beta \times C),
\end{equation}
where $m^{\ast}_{\rm{B}}$ is the observed peak magnitude, $M_{\rm{B}}$ is the absolute magnitude, $X_{1}$ is the time stretching of the light curve, and $C$ is the supernova color at maximum brightness. The luminosity distance $d_{\rm{L}}$ of a supernova in a spatially flat FRW universe is defined as
\begin{equation}
d_{{\rm L}}(z_{\rm{hel}},z)=\frac{1+z_{\rm{hel}}}{H_{0}} \int_{0}^{z} \frac{dz'}{E(z')},
\end{equation}
where $z_{\rm{cmb}}$ and $z_{\rm{hel}}$ are the CMB frame and heliocentric redshifts, respectively. The $\chi^{2}$ function for the JLA SN observation is
\begin{equation}
\chi^{2}_{\rm{SN}}=(\hat{\mu}-\mu_{\rm{th}})^{\dagger}C_{\rm SN}^{-1}(\hat{\mu}-\mu_{\rm{th}}),
\end{equation}
where $C_{\rm SN}$ is the covariance matrix of the JLA SN observation and $\mu_{\rm{th}}$ denotes the theoretical distance modulus, defined as
\begin{equation}
\mu_{\rm{th}}=5\log_{10}\frac{d_{\rm{L}}}{10\rm{pc}}.
\end{equation}

\subsection{Cosmic microwave background}

For the CMB data, we use the ``Planck distance priors" from the Planck 2015 data \cite{Ade:2015rim}. The distance priors contain the shift parameter $R$, the ``acoustic scale" $\ell_{\rm{A}}$, and the baryon density $\omega_{\rm{b}}\equiv\Omega_{\rm{b0}}h^{2}$. $R$ and $\ell_{\rm{A}}$ are defined as
\begin{equation}
R=\sqrt{\Omega_{\rm{m0}}H^2_0}(1+z_\ast)D_{\rm A}(z_\ast),
\end{equation}
\begin{equation}
\ell_{\rm A}=(1+z_\ast)\pi D_{\rm A}(z_\ast)/r_{\rm s}(z_\ast),
\end{equation}
where $\Omega_{\rm{m0}}$ is the present-day fractional energy density of matter, $D_{\rm{A}}(z_{\ast})$ is the proper angular diameter distance at the redshift of the decoupling epoch of photons $z_{\ast}$, and $r_{\rm s}(z_{\ast})$ is the comoving size of the sound horizon at $z_{\ast}$.
In a flat universe, $D_{\rm{A}}$ can be expressed as
\begin{equation}\label{DA}
D_{\rm A}(z)=\frac{1}{H_0(1+z)}\int_{0}^{z}\frac{dz'}{E(z')},
\end{equation}
and $r_{\rm s}(z)$ can be expressed as
\begin{equation}\label{39}
r_{\rm s}(z)=\frac{1} {\sqrt{3}} \int_0^{1/(1+z)} \frac{{\rm d}a}
{a^2H(a)\sqrt{1+(3\Omega_{\rm{b0}}/4\Omega_{\gamma0})a}},
\end{equation}
where $\Omega_{\rm{b0}}$ and $\Omega_{\rm{\gamma0}}$ are the present-day fractional energy densities of baryons and photons, respectively. From the measurement of CMB, we have $3\Omega_{\rm{b0}}/4\Omega_{\gamma0}=31500\Omega_{\rm b0}h^2(T_{\rm{cmb}}/2.7K)^{-4}$, where $T_{\rm cmb}=2.7255K$. The fitting formula of $z_\ast$ is given by \cite{Hu:1995en}
\begin{equation}
z_\ast=1048[1+0.00124(\Omega_{\rm{b0}}h^2)^{-0.738}][1+g_1(\Omega_{\rm{m0}}h^2)^{g_2}],
\end{equation}
where
\begin{equation}
g_1=\frac{0.0783(\Omega_{\rm{b0}}h^2)^{-0.238}}{1+39.5(\Omega_{\rm{b0}}h^2)^{0.763}},$$   $$g_2=\frac{0.560}{1+21.1(\Omega_{\rm{b0}}h^2)^{1.81}}.
\end{equation}
Using the Planck TT+LowP data, the values of the three quantities are obtained: $R=1.7488\pm0.0074$, $\ell_{\rm{A}}=301.76\pm0.14$, and $\Omega_{\rm{b}}h^{2}=0.02228\pm0.00023$. The inverse covariance matrix for them, ${\rm Cov}^{-1}_{\rm CMB}$, can be found in Ref.~\cite{Ade:2015rim},
$${\rm Cov}^{-1}_{\rm CMB}=\left(\begin{array}{ccc}
  1 & 0.54 & -0.63 \\
  0.54 & 1 & -0.43 \\
  -0.63 & -0.43 & 1
\end{array}\right).$$
The $\chi^{2}$ function for CMB is thus given by
\begin{equation}
\chi^{2}_{\rm{CMB}}=\Delta p_{i}[{\rm Cov}^{-1}_{\rm{CMB}}(p_{i},p_{j})]\Delta p_{j}, \quad \Delta p_{i}=p_{i}^{\rm{th}}-p_{i}^{\rm{obs}},
\end{equation}
where $p_{1}=\ell_{\rm{A}}$, $p_{2}=R$, and $p_{3}=\omega_{\rm{b}}$.

\subsection{Baryon acoustic oscillations}

The BAO data can be used to measure the angular diameter distance $D_{\rm{A}}(z)$
and the expansion rate of the universe $H(z)$. The BAO measurements can provide the ratio of the effective distance measure $D_{{\rm V}}(z)$ and the comoving sound horizon size $r_{\rm{s}}(z_{\rm{d}})$ for us [i.e., $\xi(z)=D_V(z)/r_s(z_d)]$. The expression of $D_{\rm{V}}(z)$ from the spherical average is
\begin{equation}
D_{\rm V}(z)=\left[(1+z)^2D^2_{\rm A}(z)\frac{z}{H(z)}\right]^{1/3},
\end{equation}
where $D_{\rm A}(z)$ is the proper angular diameter distance [see Eq.~(\ref{DA})]. $r_{\rm{s}}(z_{\rm{d}})$ is the comoving sound horizon [see Eq.~(\ref{39})] at the drag epoch with redshift $z_{\rm{d}}$. Its fitting formula is given by \cite{Eisenstein:1997ik}
\begin{equation}
z_{\rm d}=\frac{1219(\Omega_{\rm{m0}}h^2)^{0.251}}{1+0.659(\Omega_{\rm{m0}}h^2)^{0.828}}[1+b_1(\Omega_{\rm{b0}}h^2)^{b_2}],
\end{equation}
where
\begin{equation}
b_1=0.313(\Omega_{\rm{m0}}h^2)^{-0.419}[1+0.607(\Omega_{\rm{m0}}h^2)^{0.674}],
\end{equation}
\begin{equation}
b_2=0.238(\Omega_{\rm{m0}}h^2)^{0.223}.
\end{equation}

We use four BAO points from the six-degree-field galaxy survey (6dFGS) at $z_{\rm eff}=0.106$ \cite{Beutler:2011hx}, the SDSS main galaxy sample (MGS) at $z_{\rm eff}=0.15$ \cite{Ross:2014qpa}, the baryon oscillation spectroscopic survey (BOSS) ``LOWZ" at $z_{\rm eff}=0.32$ \cite{Anderson:2013zyy}, and the BOSS CMASS at $z_{\rm eff}=0.57$ \cite{Anderson:2013zyy}.
The $\chi^2$ function for BAO is given by
\begin{equation}
\chi^2_{\rm BAO}=\sum\limits_{i=1}^4 \frac{(\xi^{\rm obs}_i-\xi^{\rm th}_i)^2}{\sigma_i^2}.
\end{equation}

\subsection{The Hubble constant}

The Hubble constant direct measurement we use in this work is given by Efstathiou \cite{Efstathiou:2013via}, $H_0=70.6\pm3.3$ km s$^{-1}$ Mpc$^{-1}$. It is a re-analysis of the Cepheid data of Riess et al \cite{Riess:2011yx}. The $\chi^2$ function of the Hubble constant measurement is
\begin{equation}
\chi^2_{H_0}=\left(\frac{h-0.706}{0.033}\right)^2.
\end{equation}

\section{Results and discussion}\label{sec4}

\begin{table*}[!htp]\small
\setlength\tabcolsep{8pt}
\caption{Summary of the information criteria results.}
\label{table1}
\centering
\renewcommand{\arraystretch}{1.5}
\begin{tabular}{lccc}
\\
\hline\hline
Model  & $\chi^2_{\rm min}$ & $\Delta$AIC & $\Delta$BIC \\
  \hline

$\Lambda$CDM       & $699.3776$
                   & $0$
                   & $0$
                   \\

HDE                & $704.6058$
                   & $7.2282$
                   & $11.8456$
                   \\

IHDE1              & $699.6552$
                   & $4.2776$
                   & $13.5124$
                   \\

IHDE2              & $699.8078$
                   & $4.4302$
                   & $13.6650$
                   \\

IHDE3              & $699.7468$
                   & $4.3692$
                   & $13.6040$
                   \\

IHDE4              & $699.7050$
                   & $4.3274$
                   & $13.5622$
                   \\

IHDE5              & $699.7330$
                   & $4.3554$
                   & $13.5902$
                   \\
\hline\hline
\end{tabular}
\end{table*}

\begin{table*}[!htp]\small
\setlength\tabcolsep{5pt}
\caption{Fitting results of the models. Best-fit values with $\pm1\sigma$ errors are presented. }
\label{table2}
\renewcommand{\arraystretch}{2}\centering
\begin{tabular}{lcccccc}
\\
\hline\hline
Parameter  & HDE & IHDE1 & IHDE2 & IHDE3 & IHDE4 & IHDE5\\
  \hline
$\Omega_{\rm{m0}}$ & $0.3242^{+0.0081}_{-0.0079}$
                   & $0.3148^{+0.0084}_{-0.0103}$
                   & $0.3130^{+0.0101}_{-0.0091}$
                   & $0.3133^{+0.0101}_{-0.0093}$
                   & $0.3135^{+0.0099}_{-0.0090}$
                   & $0.3119^{+0.0117}_{-0.0080}$
                   \\

$\Omega_{\rm{b0}}$ & $0.0522^{+0.0011}_{-0.0012}$
                   & $0.0501^{+0.0031}_{-0.0035}$
                   & $0.0500^{+0.0016}_{-0.0015}$
                   & $0.0498^{+0.0018}_{-0.0015}$
                   & $0.0499^{+0.0016}_{-0.0015}$
                   & $0.0497^{+0.0019}_{-0.0014}$
                   \\

$c$                & $0.7331^{+0.0354}_{-0.0421}$
                   & $0.7133^{+0.0499}_{-0.0636}$
                   & $0.6964^{+0.0576}_{-0.0511}$
                   & $0.6933^{+0.0727}_{-0.0468}$
                   & $0.6959^{+0.0637}_{-0.0490}$
                   & $0.6895^{+0.0754}_{-0.0413}$
                   \\

$\beta$            &...
                   & $0.0340^{+0.0180}_{-0.0237}$
                   & $0.0257^{+0.0183}_{-0.0183}$
                   & $0.0134^{+0.0111}_{-0.0089}$
                   & $0.0286^{+0.0247}_{-0.0195}$
                   & $0.0558^{+0.0587}_{-0.0358}$
                   \\

$h$                & $0.6565^{+0.0076}_{-0.0068}$
                   & $0.6665^{+0.0115}_{-0.0090}$
                   & $0.6678^{+0.0106}_{-0.0113}$
                   & $0.6685^{+0.0100}_{-0.0118}$
                   & $0.6683^{+0.0101}_{-0.0115}$
                   & $0.6702^{+0.0083}_{-0.0135}$
                   \\
\hline\hline
\end{tabular}
\end{table*}

\begin{figure*}[!htp]
\includegraphics[width=11cm]{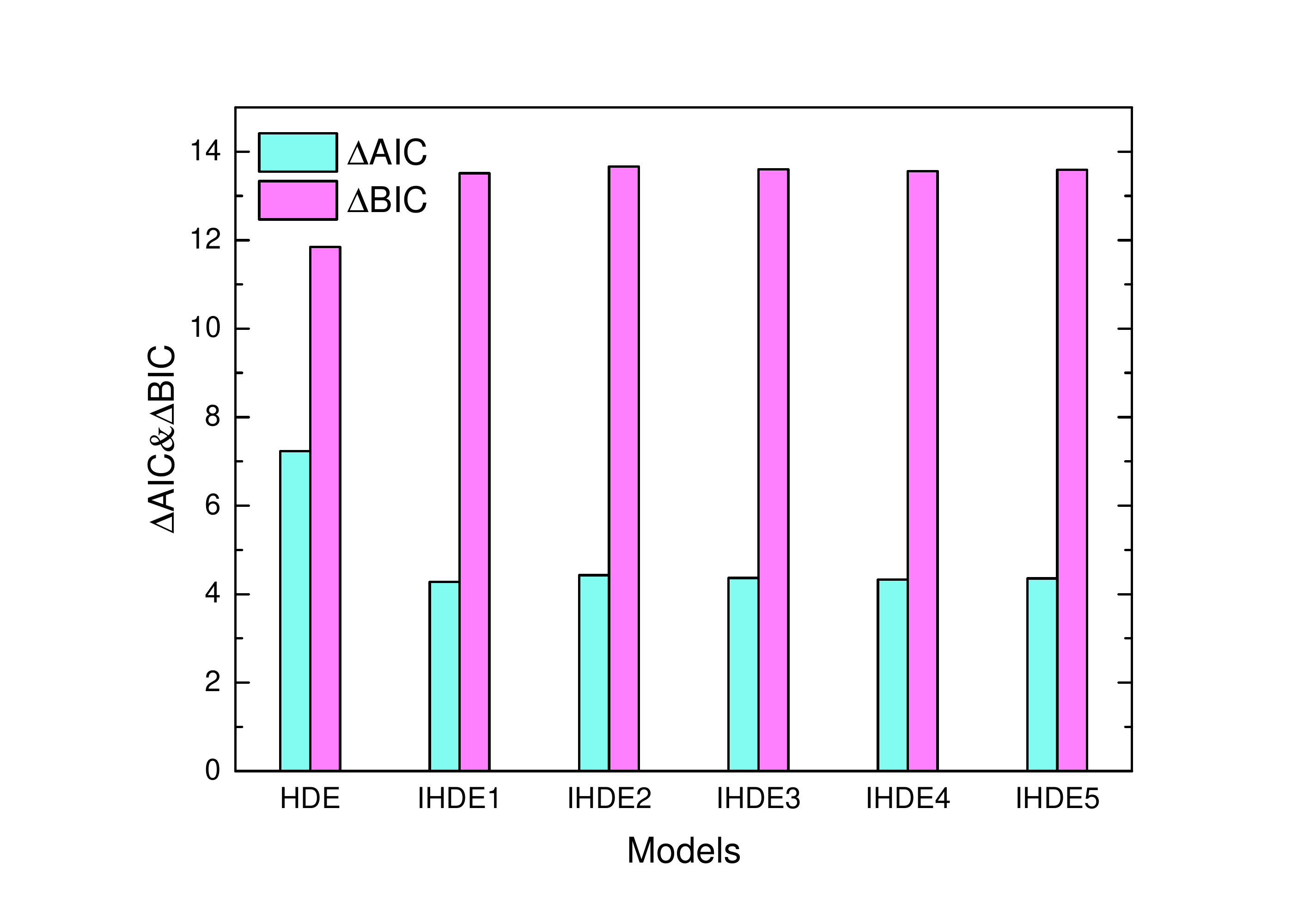}
\caption{\label{fig1}Graphical representation of the results of $\Delta$AIC and $\Delta$BIC for the HDE model and the IHDE models.}
\end{figure*}

\begin{figure*}[!htp]
\includegraphics[width=15cm]{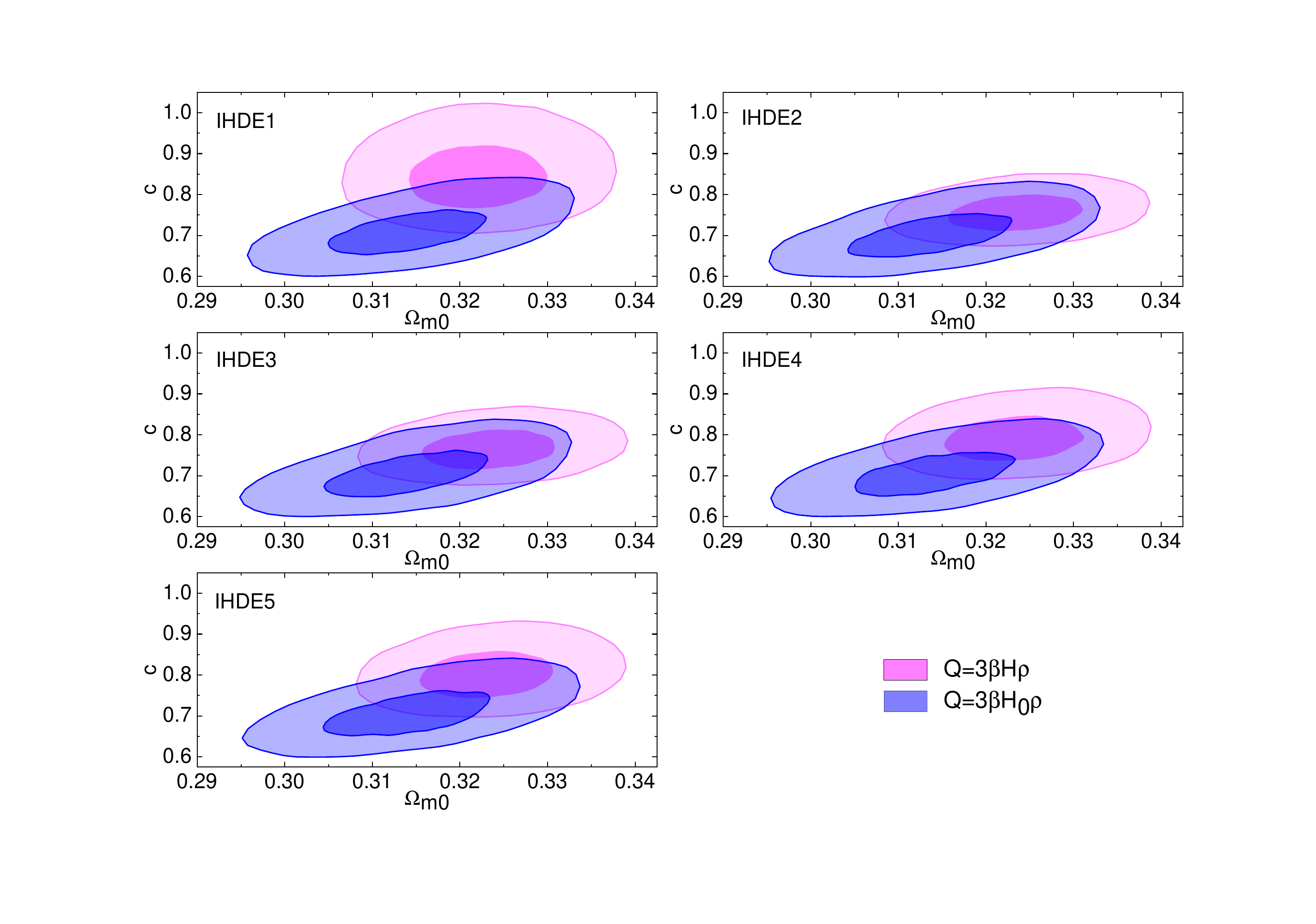}
\caption{\label{fig2}The SN+CMB+BAO+$H_0$ constraints on the HDE model and the IHDE models. The 68.3\% and 95.4\% confidence level contours are shown in the $\Omega_{\rm{m0}}$--$c$ plane. }
\end{figure*}

\begin{figure*}[!htp]
\includegraphics[width=15cm]{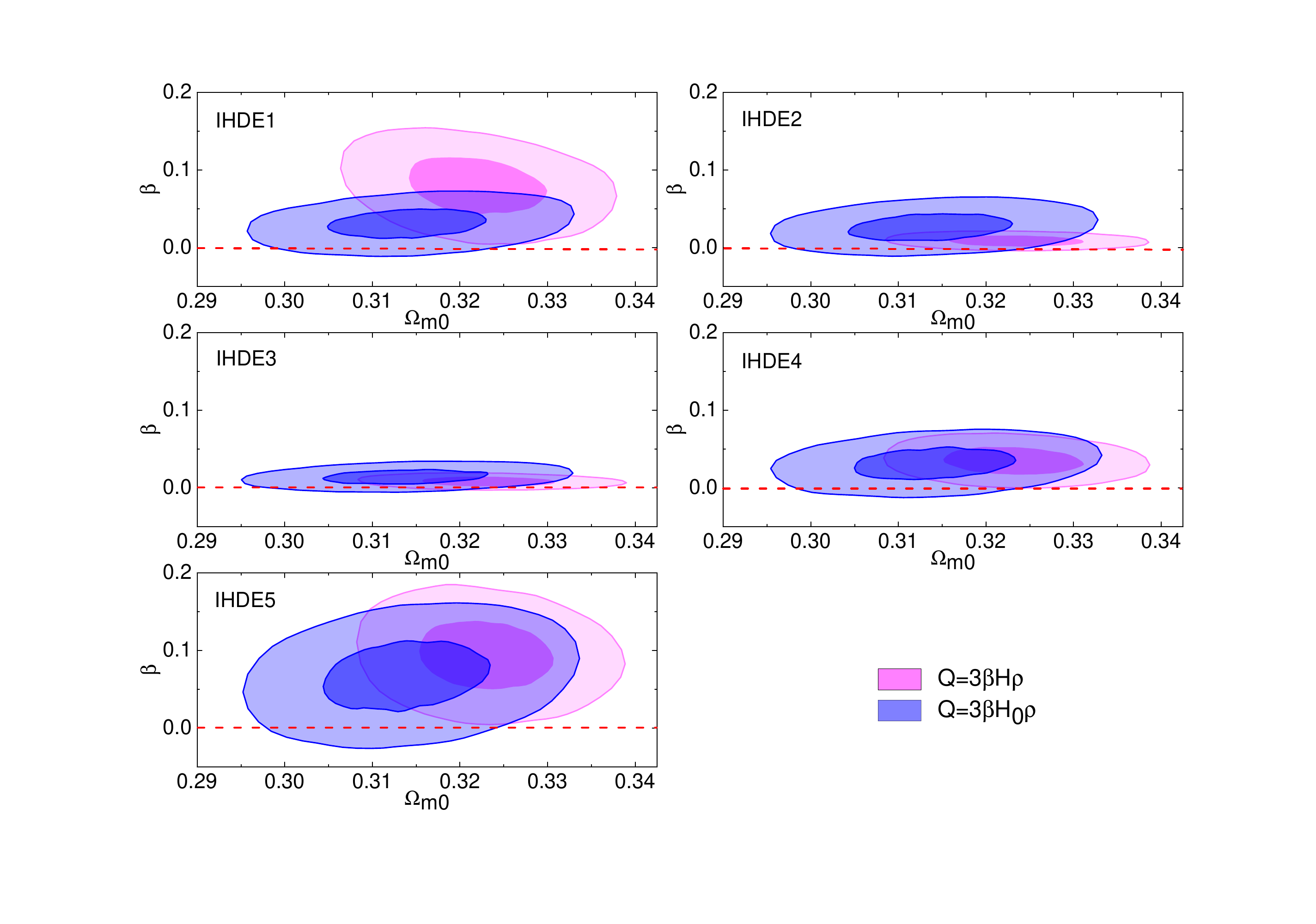}
\caption{\label{fig3}The SN+CMB+BAO+$H_0$ constraints on the HDE model and the IHDE models. The 68.3\% and 95.4\% confidence level contours are shown in the $\Omega_{\rm{m0}}$--$\beta$ plane. The red dashed line denotes the case of $\beta=0$. }
\end{figure*}

\begin{figure*}[!htp]
\includegraphics[width=15cm]{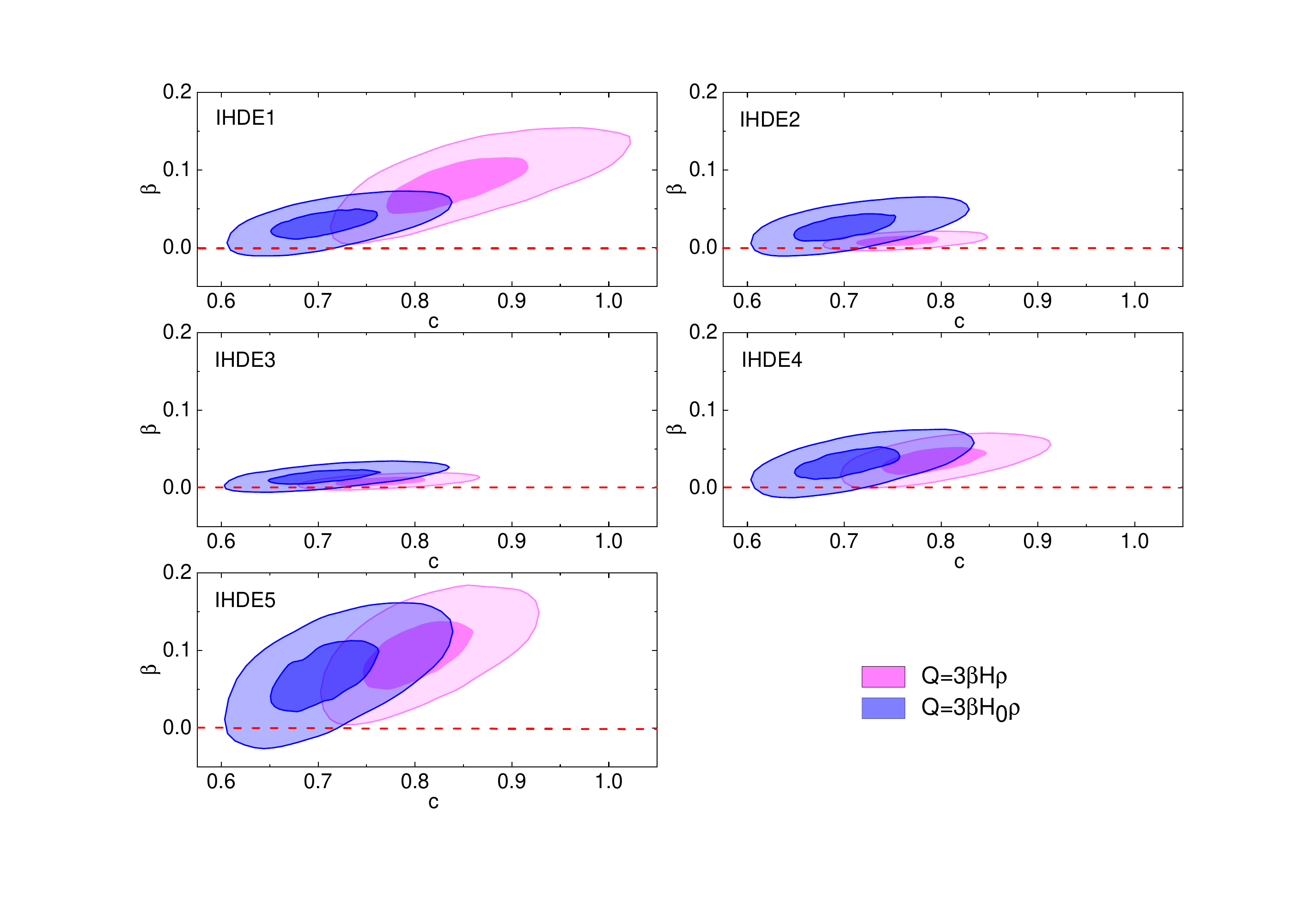}
\caption{\label{fig4}The SN+CMB+BAO+$H_0$ constraints on the IHDE models. The 68.3\% and 95.4\% confidence level contours are shown in the $c$--$\beta$ plane. The red dashed line denotes the case of $\beta=0$. }
\end{figure*}

\begin{figure*}[!htp]
\includegraphics[width=8cm]{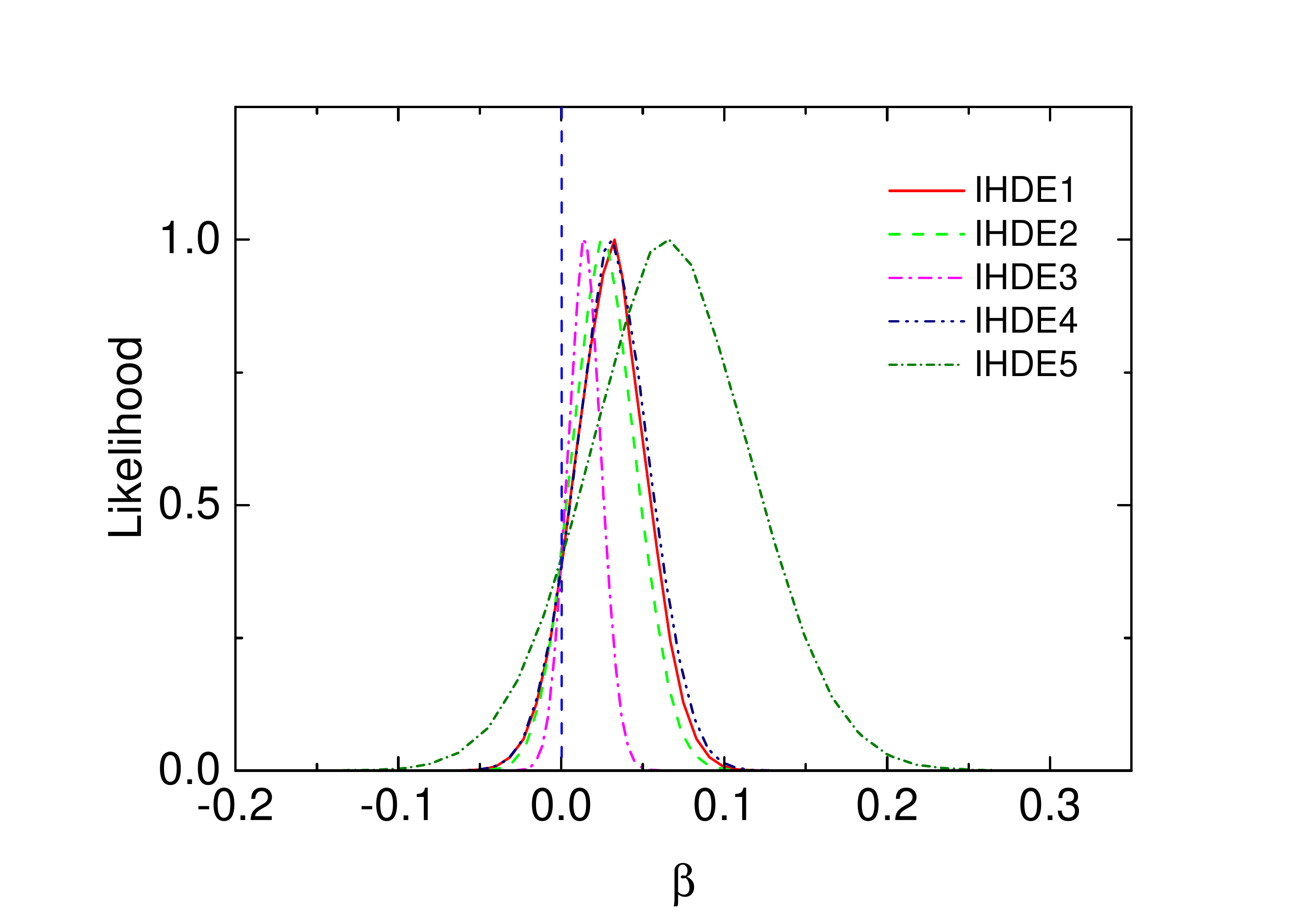}
\includegraphics[width=8cm]{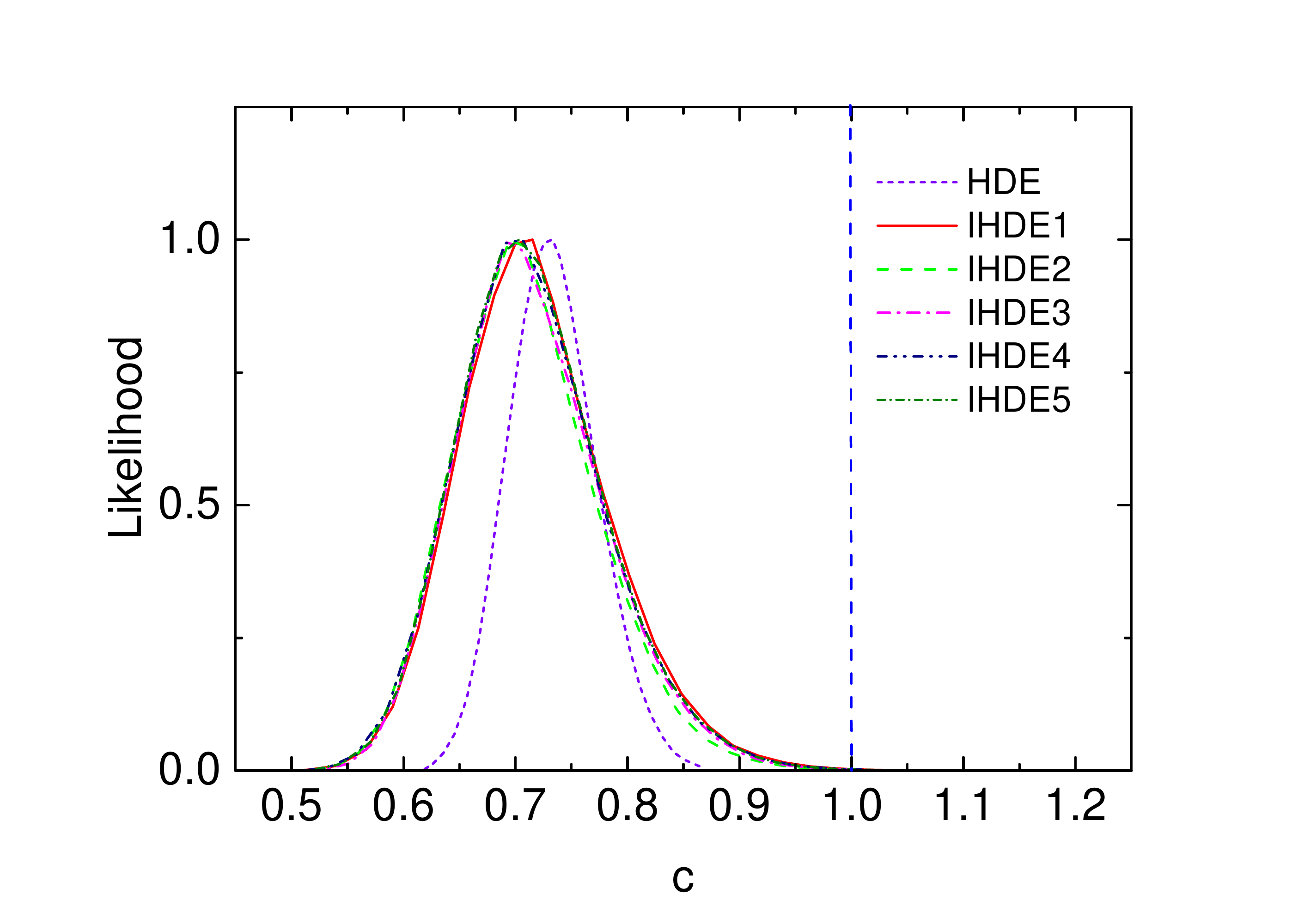}
\caption{\label{fig5}One-dimensional marginalized posterior distributions of parameters $\beta$ (left panel) and $c$ (right panel) for the HDE model and the IHDE models, from the SN+CMB+BAO+$H_0$ data. The blue dashed lines denote the cases of $\beta=0$ (left panel) and $c=1$ (right panel).}
\end{figure*}

\begin{figure*}[!htp]
\includegraphics[width=15cm]{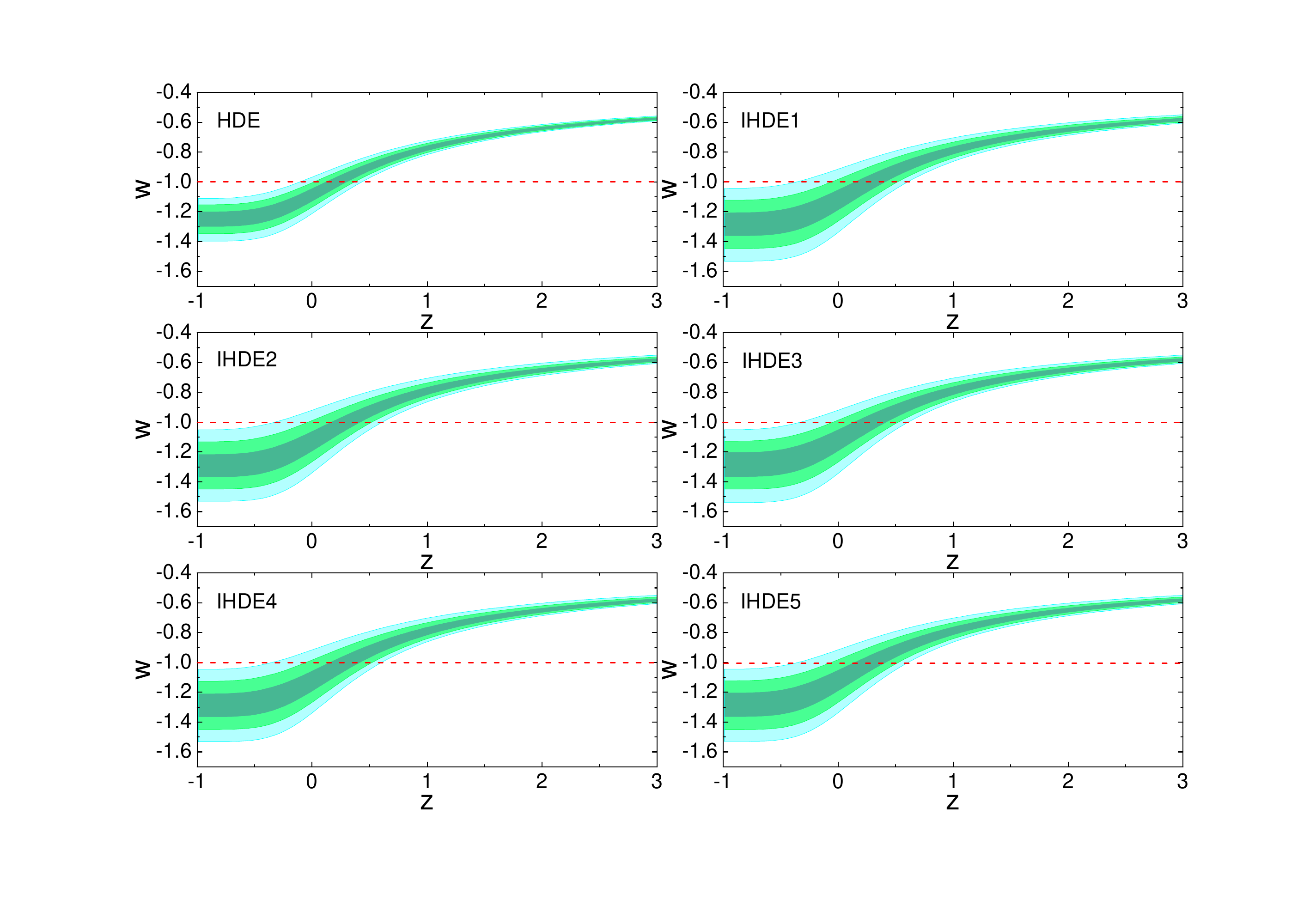}
\caption{\label{fig6}The reconstructed evolution of $w$ (with 1--3$\sigma$ errors) for the HDE model and the five IHDE models. The red dashed line denotes the cosmological constant boundary $w=-1$. }
\end{figure*}

In this section, we report the fitting results of the IHDE models and discuss the implications of these results. We use the observational data combination SN+CMB+BAO+$H_0$ to constrain the models (the $\Lambda$CDM model, the HDE model, and the IHDE1--5 models). The fitting results are summarized in Tables~\ref{table1} and \ref{table2}.

In Table~\ref{table1}, we give the values of $\chi^2_{\rm min}$, $\Delta$AIC, and $\Delta$BIC for these models. We find that, among these models, the $\Lambda$CDM model is still the best one in fitting the current observational data. The $\Lambda$CDM model has the least number of parameters, but it gets the smallest $\chi^2_{\rm min}$ value in this fit. The HDE model has one more parameter than the $\Lambda$CDM model, but it yields a greater $\chi^2_{\rm min}$ value, by $\Delta\chi^2\sim 5$. The IHDE models have two more parameters than the $\Lambda$CDM model, but they only yield similar $\chi^2_{\rm min}$ values (all around 699) to that of $\Lambda$CDM. So, although the $\Lambda$CDM model has been facing the severe theoretical problems, it is the simplest dark energy theoretical model and can explain the observations best. The HDE model indeed can provide an attractive theoretical scheme for avoiding the cosmological constant problems, but it performs worse than the $\Lambda$CDM model in fitting the observational data. It seems that one should explore more possible factors in the framework of holographic dark energy for seeking for a better model of dark energy in the sense of quantum gravity. 

Using the information criteria to assess the models (the $\Lambda$CDM model is selected as a reference model in this work), we can see that the HDE model has $\Delta {\rm AIC}=7.2$ and $\Delta {\rm BIC}=11.8$, and the IHDE1--5 models have $\Delta {\rm AIC}\sim 4.3$ and $\Delta {\rm BIC}\sim 13.6$. The comparison of HDE and IHDE shows that, only considering the factor of number of parameters (i.e., AIC), the IHDE performs better, but when further considering the factor of number of data points (i.e., BIC), the HDE performs better. We show the graphical representation of the results of $\Delta$AIC and $\Delta$BIC for the HDE model and the IHDE models in Fig.~\ref{fig1}. We also find that the values of $\chi^2_{\rm min}$ (also, $\Delta$AIC and $\Delta$BIC) for all the five IHDE models are almost equal, indicating that the current observational data equally favor these IHDE models (see also Fig.~\ref{fig1}). 

In the previous study \cite{Feng:2016djj}, Feng and Zhang investigated the IHDE models with interaction terms involving the Hubble parameter $H$ (also $Q1$--$Q5$, but with the form like $Q=3\beta H\rho$). We would like to make a comparison of our results in the present paper with those in Ref.~\cite{Feng:2016djj}. We will occasionally use the names like $Q=3\beta H_0\rho$ models and $Q=3\beta H\rho$ models to distinguish the models in this paper and those in Ref.~\cite{Feng:2016djj}. It was shown in Ref.~\cite{Feng:2016djj} that, for the $Q=3\beta H\rho$ models, according to the same data sets to this work, the IHDE5 model is the best one, the IHDE1 model is the next best one, and the IHDE2 model is the worst one. Namely, in the framework of holographic dark energy with $Q=3\beta H\rho$, the $Q=3\beta H \frac{\rho_{\rm{de}}\rho_{c}}{\rho_{\rm{de}}+\rho_{\rm c}}$ model is most favored by the current data, the $Q=3\beta H\rho_{\rm{de}}$ model is also a good model in the sense of fitting data, and the $Q=3\beta H\rho_{\rm{c}}$ model is relatively not favored by the current data. However, in the present work, we find that all the IHDE models with $Q=3\beta H_0\rho$ are equally favored by the current data.

From Table~\ref{table2}, we find that the fitting values of $c$ in both HDE and IHDE cases are all around 0.7. For the HDE model, we obtain $c=0.73$ (the best-fit value); and for the IHDE models, we obtain $c=0.69$--0.71 (the best-fit values). The $c$ value in the IHDE models is slightly smaller than that in the HDE model. For the coupling parameter $\beta$ in the IHDE models, we find that in all the cases $\beta>0$ is favored at the more than 1$\sigma$ level, indicating that the decay of dark energy into cold dark matter is favored at the more than 1$\sigma$ statistical significance by the current data. For the IHDE1, IHDE2, and IHDE4 models, we have $\beta\sim 0.03$ and $\sigma_\beta \sim 0.02$; for the IHDE3 model, we have $\beta\sim 0.01$ and $\sigma_\beta \sim 0.01$; and for the IHDE5 model, we have $\beta\sim 0.06$ and $\sigma_\beta \sim 0.04$--0.06.

In Figs.~\ref{fig2}--\ref{fig4}, we show the 1$\sigma$ and 2$\sigma$ confidence level contours in the $\Omega_{m0}$--$c$,  $\Omega_{m0}$--$\beta$, and  $c$--$\beta$ planes, respectively. The blue contours and the pink contours correspond to the $Q=3\beta H_0\rho$ models (in the present work) and the $Q=3\beta H\rho$ models (in Ref.~\cite{Feng:2016djj}), respectively. In Fig.~\ref{fig2}, we find that the $Q=3\beta H_0\rho$ models systematically move towards lower left, relative to the $Q=3\beta H\rho$ models, in the $\Omega_{m0}$--$c$ plane, indicating that for the $Q=3\beta H_0\rho$ models both $\Omega_{m0}$ and $c$ are smaller. In Figs.~\ref{fig3} and \ref{fig4}, we find that, for the IHDE1 and IHDE5 cases, the $\beta$ value in the $Q=3\beta H_0\rho$ models, relative to the $Q=3\beta H\rho$ models, is evidently smaller. In this work, we find that $\beta>0$ is favored at more than 1$\sigma$ level but less than 2$\sigma$ level, for all the cases. But for the $Q=3\beta H\rho$ models investigated in Ref.~\cite{Feng:2016djj}, it was shown that the interaction between dark energy and dark matter can be detected at more than 2$\sigma$ significance; for example, for the IHDE1 model, $\beta>0$ is favored at the 2.3$\sigma$ level, and for the IHDE5 model, $\beta>0$ is favored at the 2.1$\sigma$ level.

From Fig.~\ref{fig4}, we find that for all the IHDE models $\beta$ and $c$ are in positive correlation. A positive $\beta$ means that dark energy decays into dark matter, and thus the result of $\beta>0$ will affect the parameter estimation of $c$. The positive correlation between $\beta$ and $c$ implies that the decay of dark energy into dark matter will decrease the happening possibility of big rip in a finite future. 

In Fig.~\ref{fig5}, we show the one-dimensional marginalized posterior distributions of $\beta$ (left panel) and $c$ (right panel) for the HDE model and the IHDE models from the current observations. The blue dashed lines in the figure denote the cases of $\beta=0$ (left panel) and $c=1$ (right panel). We can see from the right panel that the $c$ values for the IHDE models are almost the same and slightly smaller than the $c$ value of the HDE model. But, although the $c$ value of the IHDE models is smaller, thanks to the positive $\beta$, the happening risk of big rip in the IHDE models, compared to the HDE model, is still decreased.(Note that, in the model of holographic dark energy, $c<1$ will lead to a late-time phantom and thus a big rip in the finite future; see, e.g., Ref.~\cite{Zhang:2005yz}.) In Fig.~\ref{fig6}, we show the reconstructed evolution of $w$ (with 1--3$\sigma$ errors) for the HDE model and the IHDE models. The red dashed line in the figure denotes the cosmological constant boundary $w=-1$. We can clearly see that, compared to the HDE model, the happening risk in the IHDE models is indeed decreased.

\section{Conclusion}\label{sec5}

In this paper, we have investigated the interacting holographic dark energy models in which the interaction term $Q$ does not involve the Hubble parameter $H$. We consider five typical IHDE models with the interaction terms $Q=3\beta H_{0}\rho_{\rm{de}}$, $Q=3\beta H_{0}\rho_{\rm{c}}$, $Q=3\beta H_{0}(\rho_{\rm{de}}+\rho_{\rm c})$, $Q=3\beta H_{0}\sqrt{\rho_{\rm{de}}\rho_{\rm c}}$, and $Q=3\beta H_{0}\frac{\rho_{\rm{de}}\rho_{c}}{\rho_{\rm{de}}+\rho_{\rm c}}$, respectively. We use the current observational data, including SN (JLA) data, CMB (Planck 2015 distance priors) data, BAO data, and $H_0$ measurement, to constrain these models.

We find that the current observational data equally favor these IHDE models. We also find that in all the cases the coupling parameter $\beta>0$ is favored at more than 1$\sigma$ level (but less than 2$\sigma$ level), indicating that the current observations slightly favor the decay of dark energy into dark matter in the current framework of IHDE (with $Q$ excluding $H$).

We have made a comparison of our results in the present work with those in the previous work \cite{Feng:2016djj} in which the IHDE models with $Q$ involving $H$ are investigated. In Ref.~\cite{Feng:2016djj}, it was shown that the IHDE5 ($Q=3\beta H\frac{\rho_{\rm{de}}\rho_{c}}{\rho_{\rm{de}}+\rho_{\rm c}}$) model and the IHDE1 ($Q=3\beta H\rho_{\rm{de}}$) model are most favored by the current data, and the IHDE2 ($Q=3\beta H\rho_{\rm{c}}$) model is relatively not favored by the current data; and, in some cases, the coupling of $\beta>0$ can be detected at more than 2$\sigma$ level (e.g., 2.3$\sigma$ in the IHDE1 model and 2.1$\sigma$ in the IHDE5 model). The comparison is shown in Figs.~\ref{fig2}--\ref{fig4}. We also show that for all the cases $\beta$ and $c$ are in positive correlation, which leads to that in the IHDE models the happening risk of big rip is decreased, compared to the HDE model.

\acknowledgments
This work was supported by the National Natural Science Foundation of China (Grants No.~11522540 and No.~11690021), the National Program for Support of Top-Notch Young Professionals, and the Provincial Department of Education of Liaoning (Grant No.~L2012087).

\end{document}